\documentclass[prl,twocolumn,tightenlines,superscriptaddress,showkeys,showpacs,lengthcheck,preprintnumbers]{revtex4}

\usepackage{graphicx}
\usepackage{amsmath}
\usepackage{amssymb}
\usepackage{dcolumn}
\newcolumntype{z}[0]{D{.}{.}{2.4}}
\newcolumntype{q}[1]{D{.}{.}{#1}}

\def\be{\begin{eqnarray}}
\def\ee{\end{eqnarray}}
\def\bea{\begin{eqnarray}}
\def\eea{\end{eqnarray}}
\def\beas{\begin{eqnarray*}}
\def\eeas{\end{eqnarray*}}
\newcommand{\eq}[1]{Eq.~(\ref{#1})}

\def\bfb{{\bf b}}

\begin{document}

\preprint{NT@UW-09-06}

\title{
 Neutron Properties in the Medium 
}
\author{I.~C.~Clo\"et}
\affiliation{Department of Physics,
University of Washington, Seattle, Washington 98195-1560}
\author{Gerald A. Miller}
\affiliation{Department of Physics,
University of Washington, Seattle, Washington 98195-1560}

\author{E. Piasetzky}
\affiliation{The Beverly and Raymond Sackler School of Exact Sciences,
Tel Aviv University, Tel Aviv 69978, Israel} 
\author{G. Ron}
\affiliation{Department of Particle Physics, Weizmann Institute of Science, Rehovot 76100, Israel}

\begin{abstract}
We demonstrate that for small values of momentum transfer, $Q^2$, the in-medium change of 
the $G_E/G_M$ form factor ratio for a bound neutron is dominated by the change in the 
electric charge radius and predict in a model independent manner that the in-medium 
ratio will increase relative to the free result. 
This effect will act to increase the predicted cross-section for the neutron recoil polarization
transfer process $^4\text{He}\left(\vec{e},e'\vec{n}\right){}^3\text{He}$.
This is in contrast to medium modification effects on the proton $G_E/G_M$ from factor ratio, 
which act to decrease the predicted cross-section for the 
$^4\text{He}\left(\vec{e},e'\vec{p}\,\right){}^3\text{H}$ reaction.
Experiments to measure the in-medium neutron
form factors via neutron knockout reactions are currently feasible in the 
range $0.1<Q^2 <1 $ GeV$^2$. 

\end{abstract}


\maketitle
The discovery by the European Muon Collaboration (EMC) that the structure function of a 
nucleus, in the valence quark region, is reduced relative to the free nucleon occurred more 
than twenty years ago \cite{EMCrefs}.
The immediate parton model interpretation is that the valence quarks in a nucleon
bound in a nucleus carry less momentum than when the nucleon is in free space.
The uncertainty principle then implies that the nucleon's size may also increase \cite{EMCrevs}.
This medium modification of nucleon structure should have consequences for nuclear 
reactions that are sensitive to the properties of a single nucleon. However unambiguous 
evidence for such modifications have not yet been observed.

Searches for medium modifications have been performed using the $(e,e')$ 
reaction \cite{eep}.
The polarization transfer reaction $(\vec{e},e'\vec{p}\,)$ 
on a proton target measures quantities  proportional to the ratio 
of the proton's electric and magnetic form factors \cite{str08}.
When such measurements are performed on a nuclear target,
\textit{e.g.}  the reaction
$^4\text{He}\left(\vec{e},e'\vec{p}\,\right){}^3\text{H}$, 
the polarization transfer observables are sensitive to the $G_E/G_M$ form factor
ratio of a  proton embedded in the nuclear environment. Several
such $^4$He experiments  have been performed \cite{polt}.
The data can be described well by including the effects of 
medium-modified form factors \cite{Lu:1997mu,Smith:2004dn,Horikawa:2005dh,ian} 
(in which the ratio is reduced by the influence of the medium)  or by
including effects from strong charge-exchange final state
interactions (FSI) \cite{Schiavilla:2004xa}.  However, the effects of the strong FSI 
may not be  consistent with measurements of the induced polarization \cite{str08}.
It is therefore important to find an alternative method to sort out the influence of  medium
modifications and FSI. The purpose of this letter is to suggest that
important progress can be achieved 
by measuring neutron recoil polarization in the 
$^4\text{He}\left(\vec{e},e'\vec{n}\right){}^3\text{He}$ reaction. 
Recent advances in experimental techniques make such considerations very timely.

Before analyzing the polarization transfer neutron knockout reaction on $^4$He,
it is worthwhile to consider the validity of the general proposition that the structure of a 
single nucleon is modified by its presence in the nuclear medium.  
The root cause of any such modification is the interaction between
nucleons, so one needs to consider whether the entire concept 
of single nucleon modification makes sense. 
Our assertion is that if the kinematics of a given experiment
select single nucleon properties,  such as in  quasi-elastic
scattering, it does make sense to
consider how a single nucleon is modified.  Thus the influence of
long-range effects, such as single pion-exchange, occur as
multi-nucleon operators and are not considered medium modifications
effects of a single nucleon, that we wish to isolate using quasi-elastic 
scattering. Within the quasi-elastic region it may be possible to characterize
these medium modifications by the virtuality of the bound nucleon 
\cite{Ciofi degli Atti:2007vx}.

We begin the analysis by considering the situation for small values of $Q^2$,
where $Q^2$ is the negative of the square of the virtual photon's four-momentum.
In this region the Sachs electric and magnetic form factors \cite{Ernst:1960zza} 
for the free proton can be expressed in the form 
\begin{align}
\label{zeroexp1}
                G_{Ep}(Q^2) &\simeq 1 - \frac{1}{6}\,Q^2\,\hat{R}_{Ep}^2,       \\
\label{zeroexp}
\frac{1}{\mu_p} G_{Mp}(Q^2) &\simeq 1 - \frac{1}{6}\,Q^2\, \hat{R}_{Mp}^2,
\end{align}
where $\mu_p$ is the proton magnetic moment and the effective electric and magnetic 
radii \cite{Miller:2007kt} are labeled by $\hat{R}_{Ep}$ and $\hat{R}_{Mp}$, respectively.
The effective radii are defined via the Sachs form factors and in a departure from the notation
of Ref.~\cite{Miller:2007kt} will be labeled with a caret, because a superscript 
$*$ is reserved  to denote in-medium quantities. 
Keeping only the leading $Q^2$ dependence, the proton electric to magnetic 
form factor ratio can be expressed as
\begin{align}
\!\mathcal{R}_p \equiv \frac{G_{Ep}(Q^2)}{G_{Mp}(Q^2)}
\simeq \frac{1}{\mu_p}\! \left[ 1-\frac{1}{6}\,Q^2(\hat{R}_{Ep}^2-\hat{R}_{Mp}^2) \right].
\label{rcp}
\end{align}
For a proton bound in a nucleus
we may define an analogous ratio which we label $\mathcal{R}_p^*$. The influence 
of the medium may change any of the three quantities $\mu_p$, $\hat{R}_{Ep}$ and $\hat{R}_{Mp}$. 
Extensive studies of the EMC effect seem to imply that the nucleon expands
in-medium. Therefore, since $\hat{R}_{Ep}^2 \simeq \hat{R}_{Mp}^2$ in free space, and if
we assume the in-medium changes are similar for the electric and magnetic radii,
the influence of the term proportional to $Q^2$ in Eq.~\eqref{rcp} would be essentially 
negligible. However, one may expect that the value of $\mu_p$ in the medium 
will increase, along with the increasing magnetic radius. 
In this scenario the super-ratio $\mathcal{R}_p^*/\mathcal{R}_p$ would be less than 
one and largely independent of $Q^2$. 

\begin{figure}[tbp]
\centering\includegraphics[width=\columnwidth]{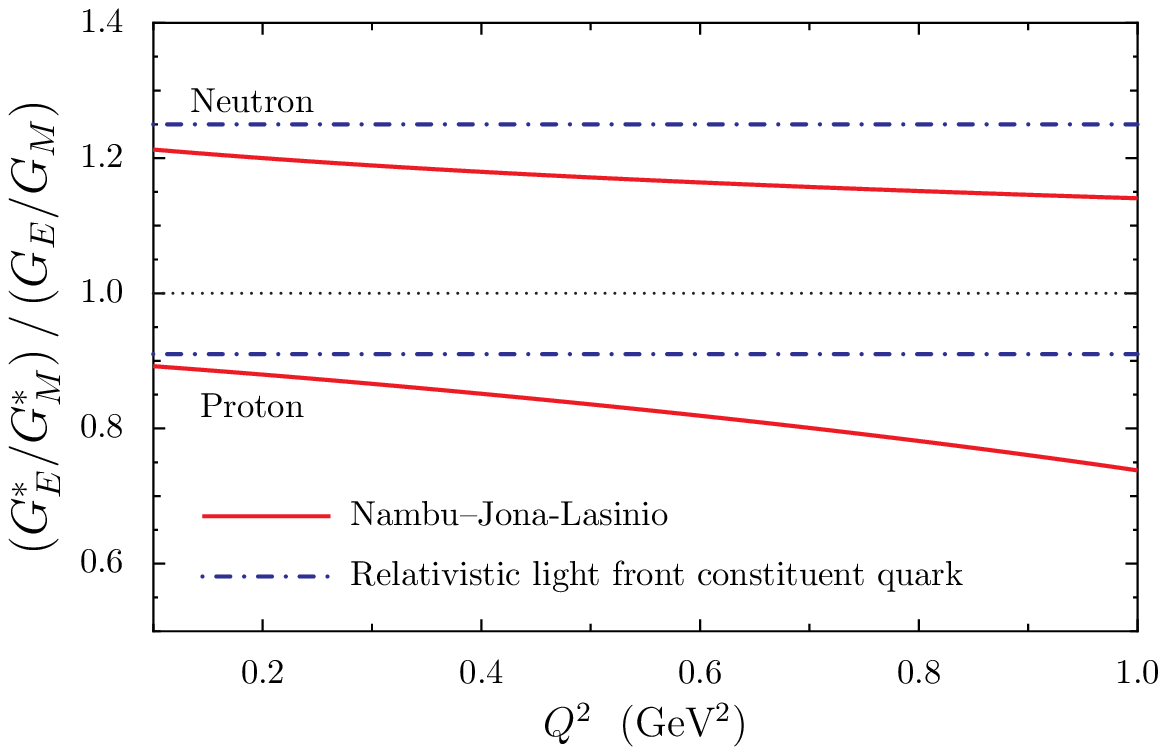}
\vspace{-2em}
\caption{Super-ratios for the proton and neutron form factors in nuclear matter,
obtained from the Nambu--Jona-Lasinio model of Ref.~\cite{ian}
and the relativistic light front constituent quark model of Ref.~\cite{Chung:1991st}.
\vspace{-1em}} 
\label{fig:2}
\end{figure}

This expectation is borne out by specific model calculations
\cite{Lu:1997mu,Horikawa:2005dh,ian} and, more importantly, by the experimental data in Refs.~\cite{polt}.
The basic idea behind the models is that confined quarks in a nucleon -- which is treated as a MIT bag 
in Ref.~\cite{Lu:1997mu} or as a solution of the relativistic Faddeev equation in Refs.~\cite{Horikawa:2005dh,ian} --
are influenced by the quarks of neighboring nucleons through the
exchange of a scalar meson, which provides the necessary attraction to bind the nucleus. 
The results of Ref.~\cite{ian} for the proton super-ratio in nuclear matter are given in Fig.~\ref{fig:2}.
A contrasting model is that of Smith \& Miller \cite{smith,Smith:2004dn}, where
the quarks are confined in a chiral soliton which is identified as the nucleon.
In-medium the confined quarks are also influenced by the exchange of scalar 
objects between quarks of neighboring nucleons.
In this model the magnetic properties are dominated by the sea, which is resistant to
the influence of the medium. Thus $\mu_p$ and $\hat{R}_{Mp}$ remain largely unchanged
whereas $\hat{R}_{Ep}$ increases. Once again the super-ratio $\mathcal{R}_p^*/\mathcal{R}_p$ is
less than unity, however in this model it is expected to vary linearly with $Q^2$.
As noted earlier, in the region where data exist all three models are in
satisfactory agreement with experiment, as are  
the effects of including FSI. 

There are two lessons from this. Firstly, very different models predict the super-ratio 
to be less than one for the proton, but for very different reasons. Thus  
there is a need for another experimental way to determine which, if
any, of the relevant parameters are changed in the medium. Secondly, we need more
precise data and an increase in the $Q^2$ range of the $(\vec{e},e'\vec{p}\,)$ experiments.

One way to help resolve the different mechanisms responsible for the medium modification
of nucleons and to also determine the influence of FSI  is to consider the
neutron in the medium. The analogous expression to \eq{rcp} for the neutron,
valid at small $Q^2$, is
\begin{align}
\mathcal{R}_n \equiv \frac{G_{En}(Q^2)}{G_{Mn}(Q^2)} \simeq 
- \frac{1}{\mu_n}\,\frac{1}{6}\,Q^2\,\hat{R}_{En}^2,
\label{freen}
\end{align}
where the effective magnetic radius does not appear, since it is the coefficient of
a $Q^4$ term. We immediately see that, in contrast with the proton,
the  medium modifications are generally expected to depend on
possible changes in both the electric radius and magnetic moment.
This implies that the behavior of the super-ratio $\mathcal{R}_n^*/\mathcal{R}_n$ 
at small $Q^2$ is determined by a
competition between the expected increases in both these quantities.
The electric radius is more important in \eq{freen} because it enters
quadratically. Thus one may expect, in contrast with the proton, that the neutron
super-ratio will be larger than one.

It is worthwhile to consider specific models as examples of the
previous general statements. In the quark-diquark Nambu--Jona-Lasinio (NJL) 
model of Refs.~\cite{Horikawa:2005dh,ian}, both $\hat{R}_{En}$ and $\hat{R}_{Mn}$ increase in-medium, 
however there is only a small in-medium change in the neutron magnetic moment.
Therefore at low $Q^2$ one finds that the super-ratio is dominated by the
change in $\hat{R}_{En}$ and therefore increases. This is shown in Fig.~\ref{fig:2}
where the results of Ref.~\cite{ian} are illustrated.
In the model of Smith \& Miller \cite{smith} the value of $\mu_n$ and $\hat{R}_{Mn}$
are largely unchanged in the medium, however $\hat{R}_{En}$ increases. 
Therefore both models predict that the super-ratio goes up for the neutron and
down for the proton.

We can also consider placing the relativistic light front constituent quark model
of Ref.~\cite{Chung:1991st} in the medium. 
This model for the nucleon is characterized in free space by a confinement
scale $1/\alpha$ and a quark mass $m_q$. One might imagine that the medium
changes each of these quantities. Numerically the change in $m_q$ 
is more important for the magnetic moments, and we find that in the medium
$\delta \kappa_p/\kappa_p\approx \delta \kappa_n/\kappa_n\approx -\delta m_q/m_q$. 
Thus in this model the percentage change in the neutron and proton 
anomalous magnetic moments is the same. The nucleon charge and magnetic radii
are proportional to $1/\alpha$, therefore the percentage in-medium change of
the radii behaves like $\sim \delta\alpha/\alpha$. For the proton, the super-ratio 
is therefore dominated by the change in the magnetic moment, see Eq.~\eqref{rcp}, so that the prediction of
this model is that the proton super-ratio is less than unity. For the neutron 
the change in the radius enters quadratically, see Eq.~\eqref{freen}, 
so that once again one expects an increase in the neutron super-ratio.
We estimate the size of these in-medium effects by using Eqs.~\eqref{rcp} and \eqref{freen}, along 
with the appropriate expressions in Ref.~\cite{Chung:1991st}. Assuming medium effects
increase the radii by 10\% and reduce the quark mass by 20\% we obtain the 
results illustrated in Fig.~\ref{fig:2}.

It is possible to generalize our arguments so that they are applicable
beyond the $Q^2$ domain where Eqs.~\eqref{rcp} and \eqref{freen} hold, 
namely $\tfrac{1}{6}\,Q^2\,\hat{R}^2_{Ep,Mp} \ll 1$. The current $Q^2$ range where the
neutron polarization transfer reaction is experimentally feasible is probably
between $0.1$ and $1$ GeV$^2$. In this region the proton electromagnetic form factors 
fall monotonically, so that one may characterize the size of a system by thinking
of the width of the electromagnetic form factors $G_{Ep,Mp}(Q^2)$ 
as a measure of the inverse of the square of a generalized radius. 
The expectations about the influence of the medium on the generalized radii would be
essentially the same as for the radii of Eqs.~\eqref{zeroexp1} and \eqref{zeroexp}. 
The concept of generalized radii for the neutron is potentially even
more interesting, because $G_{En}$ rises from zero at $Q^2=0$ to a peak at about 
$Q^2=0.4\,$GeV$^2$ and then falls monotonically, whereas  $G_{Mn}(Q^2)$ simply falls monotonically. 
One may  then characterize the square of a generalized radius in terms of the value
of $Q^2$ for which $G_{En}$ peaks. This generalized radius can be
expected to increase in the medium.

It is intriguing that each of the three models described earlier find 
that $\mathcal{R}^*/\mathcal{R}$ is greater than unity for neutrons
and less than unity for protons. We shall now try to understand this
from a more formal perspective, for both 
 magnetic moments and radii. 
Consider the expression for the anomalous magnetic moment $\kappa$ derived in 
Ref.~\cite{Miller:2007kt}, namely
\begin{align}
\kappa = \langle X\vert\sum_q e_q\int d^2b \;b_y\;{q}_+^\dagger(0,\bfb)q_+(0,\bfb)\vert X\rangle,
\label{mua}  
\end{align}
where  $q_+(x^-,\bfb)$ is a quark-field operator of charge $q$ and $\textbf{b}$ is the impact
parameter. The subscript $+$ indicates a lightcone good component of the quark field, 
defined by $q_+=\gamma^0\gamma^+q$, and therefore the operator ${q}_+^\dagger(0,\bfb)q_+(0,\bfb)$
is a number operator for valence quarks with impact parameter $\textbf{b}$.
Explicitly the state $\vert X\rangle$ has the form
\begin{align}
\vert X\rangle &\equiv \frac{1}{\sqrt{2}}\left[\vert X, +\rangle + \vert X, -\rangle\right] \nonumber \\
&\equiv \frac{1}{\sqrt{2}}\left[\left|p^+,{\bf R}= {\bf 0},
+\right\rangle+\left|p^+,{\bf R}= {\bf 0},
-\right\rangle\right],
\label{eq:statex}
\end{align}
where the first term in Eq.~\eqref{eq:statex} represents a transversely localized 
state of definite $p^+$ momentum and positive light-cone helicity, whereas the second
state has negative light-cone helicity.
The state $\vert X\rangle$ may be interpreted as that of
a transversely polarized target \cite{Burkardt:2002ks,Burkardt:2002hr}, up to relativistic corrections caused by
the transverse localization of the wave packet \cite{Burkardt:2005hp}.

Define the contribution of the $u$-quarks to the proton matrix element
as $2u$, where
\begin{align}
2u = \langle X\vert\int d^2b \;b_y\;{{q}_u}_+^\dagger(0,\bfb){q_u}_+(0,\bfb)\vert X\rangle,
\end{align}
and the contribution of the $d$-quarks as
\begin{align}
d = \langle X\vert\int d^2b \;b_y\;{q_d}_+^\dagger(0,\bfb){q_d}_+(0,\bfb)\vert X\rangle.
\end{align}
With this definition, and neglecting the contribution from heavy quark flavours,
the proton anomalous magnetic moment can be expressed as
\begin{align}
\kappa_p = \frac{4}{3}\,u - \frac{1}{3}\,d.  
\end{align}
Then assuming charge symmetry \cite{csb}, so that the $u$- and $d$-quark contributions 
in the proton equal the $d$- and $u$-quark contributions in the neutron, we obtain
\begin{align}
\kappa_n = -\frac{2}{3}\,u + \frac{2}{3}\,d.  
\end{align}
In the medium the nucleon matrix elements are modified. Thus $u $ and
$d$ are shifted from their free values by $\delta u$ and $\delta d$
respectively. We see  no general, model-independent way to relate
these two quantities, even in the case of symmetric nuclear
matter (with $N=Z$) where the external forces on the confined quarks
are flavor independent. This is because of the necessary interplay
between  the quark orbital angular momentum and spin.
Thus the changes in the anomalous magnetic moments are simply
\begin{align}
\delta \kappa_p &= ~~~\frac{4}{3}\,\delta u - \frac{1}{3}\,\delta d,       \\
\delta \kappa_n &= -\frac{2}{3}\,\delta u + \frac{2}{3}\,\delta d. 
\end{align}
To determine each of these quantities requires a measurement of both the 
proton and neutron magnetic moment in the medium.
An important point that is worth highlighting is that the change in the proton
magnetic moment does not simply equal $\delta \kappa_p$. If the mass of the 
nucleon changes in the medium there is also a contribution from the Dirac
form factor. If the proton magnetic moment is expressed in nuclear 
magnetons, its change in-medium is given by $\delta \kappa_p$ plus
the term $M/M^*-1$, where $M$ is the free nucleon mass and $M^*$ is the 
in-medium mass shifted by the influence of the nuclear binding potentials.

Using the relation that the transverse charge density is the two-dimensional 
Fourier transform of $F_1$ \cite{Soper:1976jc,Burkardt:2002hr,diehl2,MillerNeutron,Carlson:2007xd},
one may analyze the nucleon radii in a similar fashion to the anomalous magnetic 
moments. The $u$-quark sector contribution to the $F_1$ electric charge radius squared
is given by
\begin{align}
2R_{1u}^2 = \langle X,+ \vert 
\int d^2b\,\frac{3}{2}\,b^2{{q}_u}_+^\dagger(0,\bfb){q_u}_+(0,\bfb)\vert X,+ \rangle,
\end{align}
and the contribution from the $d$-quarks is
\begin{align}
R_{1d}^2 = \langle X,+ \vert 
\int d^2b\,\frac{3}{2}\,b^2{{q}_d}_+^\dagger(0,\bfb){q_d}_+(0,\bfb)\vert X,+ \rangle.
\end{align}
The factor 3/2 accounts for the two-dimensional integration. Recalling that
$G_E = F_1 - \tfrac{Q^2}{4\,M^2}F_2$, the effective charge radii related to $G_E$
are given by
\begin{align}
\hat{R}_{Ep}^2 &= ~~~\frac{4}{3}\, R_{1u}^2 - \frac{1}{3}\,R_{1d}^2 + \frac{3}{2\,M^2}\kappa_p,  \\
\hat{R}_{En}^2 &=   -\frac{2}{3}\, R_{1u}^2 + \frac{2}{3}\,R_{1d}^2 + \frac{3}{2\,M^2}\kappa_n.
\end{align}
For the neutron (but not the proton) the Foldy term \cite{Foldy:1952wa}, $\tfrac{3}{2M^2}\,\kappa_n$, is 
by far the dominant contribution to the charge radius. In-medium
this will almost certainly remain true. Therefore for small values of $Q^2$ 
the leading term of the neutron super-ratio is given by
\begin{align}
\frac{\mathcal{R}_n^*}{\mathcal{R}_n} \simeq \left(\frac{M}{~M^*}\right)^2,
\label{eq:neutronsuper}
\end{align}
because the anomalous magnetic moments in the Foldy terms cancel the neutron magnetic 
moments. Binding effects imply that $M^* < M$ and therefore we 
have obtained on general grounds that at small 
$Q^2$ the super-ratio should be greater than one 
for the neutron. 

This general prediction is worthy of an experimental test, and
recent technical developments make this an ideal time to plan such an experiment. 
Using recoil polarization, high precision, low $Q^2$ measurements of the 
free proton \cite{Ron:2007vr} and neutron \cite{Madey:2003av} form factors
have already been performed. With a straightforward extension of these experiments,
using a similar experimental setup, at for example, the Thomas Jefferson National 
Accelerator Facility (JLab), it would be possible to perform low $Q^2$ measurements
of the reactions $p(\vec{e},e'\vec{p}\,)$, $d(\vec{e},e'\vec{p}\,)n$, $d(\vec{e},e'\vec{n})p$, 
$^4\text{He}(\vec{e},e'\vec{p}\,)^3\text{H}$ and $^4\text{He}(\vec{e},e'\vec{n})^3\text{He}$.
This would allow a  direct test of the predictions made in this letter.
Because of the large cross-section for these reactions at low $Q^2$, and the availability
of a high current polarization and duty factor electron beam at JLab, these experiments would 
achieve excellent statistical precision within a relatively short time period. 
Such experiments could also probe the $Q^2$ dependence of the form factor  super-ratios. 
An experimental proposal to this effect is being developed by the authors for the JLab facility.

Understanding how a nucleon is modified when in the nuclear environment remains
a central challenge for the nuclear physics community. In this letter we present
a unique model independent result pertaining to the structure of a bound nucleon, 
which is expressed in Eq.~\eqref{eq:neutronsuper}, and states that the neutron super-ratio
is greater than one at small $Q^2$.
We therefore conclude that the measurement of $(\vec{e},e'\vec{n})$ processes 
on nuclear targets can provide important additional and complementary information to that already
obtained using the $(\vec{e},e'\vec{p}\,)$ reaction.
These measurements would provide an independent test of any model seeking to explain the
EMC effect and offer the hope of providing its long-sought universally
accepted explanation.


We thank S. Strauch for helpful discussions. We also thank the USDOE (FG02-97ER41014) and   
the Israel Science Foundation for partial support of this work.


\begin{thebibliography}{100}

\bibitem{EMCrefs}
 \vspace{-1em}
  J.~J.~Aubert {\it et al.}  [European Muon Collaboration],
  Phys.\ Lett.\  B {\bf 123}, 275 (1983);
  R.~G.~Arnold {\it et al.},
  Phys.\ Rev.\ Lett.\  {\bf 52}, 727 (1984);
  A.~Bodek {\it et al.},
  Phys.\ Rev.\ Lett.\  {\bf 51}, 534 (1983).

\bibitem{EMCrevs} For example, see the reviews:
  M.~Arneodo,
  Phys.\ Rept.\  {\bf 240}, 301 (1994);
  D.~F.~Geesaman, K.~Saito and A.~W.~Thomas,
  Ann.\ Rev.\ Nucl.\ Part.\ Sci.\  {\bf 45}, 337 (1995);
  G.~Piller and W.~Weise,
  Phys.\ Rept.\  {\bf 330}, 1 (2000).

\bibitem{eep}
I.~Sick, D.~Day and J.~S.~Mccarthy,
  Phys.\ Rev.\ Lett.\  {\bf 45}, 871 (1980);
J.~Jourdan,
  Phys.\ Lett.\  B {\bf 353}, 189 (1995);
 J.~Morgenstern and Z.~E.~Meziani,
  Phys.\ Lett.\  B {\bf 515}, 269 (2001).

\bibitem{str08}
S.~Malace, M.~Paolone and S.~Strauch  [Jefferson Lab Hall A Collaboration],
  AIP Conf.\ Proc.\  {\bf 1056}, 141 (2008).

\bibitem{polt}  S.~Dieterich {\it et al.},
  Phys.\ Lett.\  B {\bf 500}, 47 (2001);
  S.~Strauch {\it et al.}  [Jefferson Lab E93-049 Collaboration],
  Phys.\ Rev.\ Lett.\  {\bf 91}, 052301 (2003);
  Jefferson Lab experiment E03-104, R.~Ent, R.~Ransome, S.~Strauch and
  P.~Ulmer (spokespeople).

\bibitem{Lu:1997mu}
  D.~H.~Lu, A.~W.~Thomas, K.~Tsushima, A.~G.~Williams and K.~Saito,
  Phys.\ Lett.\  B {\bf 417}, 217 (1998);
  D.~H.~Lu, K.~Tsushima, A.~W.~Thomas, A.~G.~Williams and K.~Saito,
  Phys.\ Rev.\  C {\bf 60}, 068201 (1999).

\bibitem{Smith:2004dn}
 J.~R.~Smith and G.~A.~Miller,
  Phys.\ Rev.\ Lett.\  {\bf 91}, 212301 (2003)
  [Erratum-ibid.\  {\bf 98}, 099902 (2007)];
  J.~R.~Smith and G.~A.~Miller,
  Phys.\ Rev.\  C {\bf 70}, 065205 (2004);
P.~Lava, J.~Ryckebusch and B.~Van Overmeire,
  Prog.\ Part.\ Nucl.\ Phys.\  {\bf 55}, 437 (2005);
G.~A.~Miller,
  Eur.\ Phys.\ J.\  A {\bf 31}, 578 (2007).

\bibitem{Horikawa:2005dh}
  T.~Horikawa and W.~Bentz,
  Nucl.\ Phys.\  A {\bf 762}, 102 (2005).

\bibitem{ian} 
  I. C. Clo\"et, W.~Bentz, A.~W.~Thomas, forthcoming publication.

\bibitem{Schiavilla:2004xa}
  R.~Schiavilla, O.~Benhar, A.~Kievsky, L.~E.~Marcucci and M.~Viviani,
  Phys.\ Rev.\ Lett.\  {\bf 94}, 072303 (2005).

\bibitem{Ciofi degli Atti:2007vx}
  C.~Ciofi degli Atti, L.~L.~Frankfurt, L.~P.~Kaptari and M.~I.~Strikman,
  Phys.\ Rev.\  C {\bf 76}, 055206 (2007).

\bibitem{Ernst:1960zza}
  F.~J.~Ernst, R.~G.~Sachs and K.~C.~Wali,
  Phys.\ Rev.\  {\bf 119}, 1105 (1960).

\bibitem{Miller:2007kt}
  G.~A.~Miller, E.~Piasetzky and G.~Ron,
  Phys.\ Rev.\ Lett.\  {\bf 101}, 082002 (2008).

\bibitem{smith} 
  J.~R.~Smith and G.~A.~Miller,
  Phys.\ Rev.\ Lett.\  {\bf 91}, 212301 (2003)
  [Erratum-ibid.\  {\bf 98}, 099902 (2007)].

\bibitem{Chung:1991st}
  P.~L.~Chung and F.~Coester,
  Phys.\ Rev.\  D {\bf 44}, 229 (1991).

\bibitem{Burkardt:2002hr}
  M.~Burkardt,
  Int.\ J.\ Mod.\ Phys.\  A {\bf 18}, 173 (2003).

\bibitem{Burkardt:2002ks}
  M.~Burkardt,
  Phys.\ Rev.\  D {\bf 66}, 114005 (2002).

\bibitem{Burkardt:2005hp}
  M.~Burkardt,
  Phys.\ Rev.\  D {\bf 72}, 094020 (2005).

\bibitem{csb}
 G.~A.~Miller, B.~M.~K.~Nefkens and I.~Slaus,
  Phys.\ Rept.\  {\bf 194}, 1 (1990);
 G.~A.~Miller,
  Phys.\ Rev.\  C {\bf 57}, 1492 (1998);
G.~A.~Miller, A.~K.~Opper and E.~J.~Stephenson,
  Ann.\ Rev.\ Nucl.\ Part.\ Sci.\  {\bf 56}, 253 (2006).

\bibitem{Soper:1976jc}
  D.~E.~Soper,
  Phys.\ Rev.\  D {\bf 15}, 1141 (1977).

\bibitem{diehl2} 
  M.~Diehl,
  Eur.\ Phys.\ J.\  C {\bf 25}, 223 (2002)
  [Erratum-ibid.\  C {\bf 31}, 277 (2003)].

\bibitem{MillerNeutron} G.~A.~Miller,
  Phys.\ Rev.\ Lett.\  {\bf 99}, 112001 (2007).

\bibitem{Carlson:2007xd}
 C.~E.~Carlson and M.~Vanderhaeghen,
  Phys.\ Rev.\ Lett.\  {\bf 100}, 032004 (2008).

\bibitem{Foldy:1952wa}
  L.~L.~Foldy,
  Phys.\ Rev.\  {\bf 87}, 688 (1952).

\bibitem{Ron:2007vr}
  G.~Ron {\it et al.},
  Phys.\ Rev.\ Lett.\  {\bf 99}, 202002 (2007).

\bibitem{Madey:2003av}
  R.~Madey {\it et al.}  [E93-038 Collaboration],
  Phys.\ Rev.\ Lett.\  {\bf 91}, 122002 (2003).

\end{thebibliography}
\end{document}